\documentclass[useAMS,usenatbib]{mn2e}
\usepackage{graphicx}
\newcommand{\hangpar}{\noindent\hangindent.1in}
\newcommand{\teff}[1]{$T_{\rm eff}$}
\newcommand{\vsini}[1]{$v\cdot\sin(i)$}
\def\GR{\cal \char'122}

\title[HR 710 (HD 15144): An ultra-Sr-Rich, magnetic Ap star]
{HR 710 (HD 15144): An ultra-Sr-rich, magnetic Ap star with a close
companion\thanks{Based on observations
obtained at the European Southern
Observatory, Paranal and La Silla, Chile (ESO programmes 076.D-0169(A) and 076.C-0172(A)).}}
\author[C. R. Cowley, and S. Hubrig]
{C. R. Cowley${^1}$,
\thanks{E-mail: cowley@umich.edu\,(CC); shubrig@eso.org\,(SH)}
 S. Hubrig${^2}$   \\
$^{1}$Department of Astronomy, University of Michigan,
   Ann Arbor, MI 48109-1042, USA\\
$^{2}$European Southern Observatory, Casilla 19001,
Santiago 19, Chile\\
}
\begin{document}

\date{Accepted . Received ; in original form }

\pagerange{\pageref{firstpage}--\pageref{lastpage}} \pubyear{2007}

\maketitle

\label{firstpage}

\begin{abstract}
The magnetic, chemically peculiar A-star HR 710 (HD 15144,
ADS 1849A, AB Cet) has a close companion in a nearly
circular orbit.  Its 3-day period is unusually short
for such stars.
The system emits moderately hard x-rays, which likely
come from a white dwarf secondary (ADS 1849Ab).
The Sr\,{\sc ii} spectrum
is very strong, and the resonance lines show similar core-nib
structure to the stronger Ca K line.
We place only loose constraints on a model.  There are
indications of a lower electron/gas pressure than
expected from the star's parallax and brightness.
Strong-line profiles and anomalous excitation/ionization
indicate significant deviations
from traditional models, fixed by $T_{eff}$, $\log{g}$,
and abundances.  Weak spectral-line profiles and wavelength
shifts probably indicate abundance patches rather than
the presence of a secondary (ADS 1849Ab) spectrum.
Our UVES spectra are from only one epoch.  Additional
low-noise, high-resolution spectra are needed.
We discuss the spectrum
within the context of abundance stratification.

\end{abstract}


\section{Introduction}

Roman (1949) listed HR 710 (HD 15144) among the probable members
of the Ursa Major Stream.  She noted the strengths of
$\lambda\lambda$4077 and 4215.  More recently, King,
et al. (2003) enter a question mark in the column
for ``Final Memb.'' for HD 15144A.
Tokovinin (1997) concludes that the system ADS 1849AB is
a physical double, rather than a common proper motion pair,
but the period must be very long.  The primary, ADS 1849A (or
is itself a spectroscopic binary (Aa and Ab)
with a well-determined period
of 2.998 days (Bonsack 1981, Tokovinin 1997).
This a very short period for a magnetic Ap star.
One system, HD 200405 (cf. Carrier,
et al. 2002), with an even shorter period
is known--1.64 days.
The HR 710 system appears to be
single lined, so there has been a large uncertainty in
the mass and luminosity of the companion.

H\"{u}nsch, Schmitt, and Voges (1998, henceforth HSV)
list HR 710 in their {\em ROSAT} survey of main sequence
and subgiant stars of the Bright Star Catalogue.  They
found an x-ray
luminosity of 4.2$\times 10^{29}$ erg s$^{-1}$.  This is among
the top third of the group.  The hardness
ratio they report, 0.12, is in the upper 15\%.  Thus
HR 710 is significant, though not outstanding among
these x-ray emitters.

Czesla and Schmitt (2007) report {\em Chandra} observations
of HR 710 as well as the companion ADS 1849B in the
energy band 0.5 to 2\,keV
Their x-ray flux for the primary is in reasonable
agreement with the {\em ROSAT} data.

Extensive magnetic field measurements were made
by Babcock (1958), and Bonsack (1981).  The field
varied between $-$0.45 and $-$1.10 kG---it did
not reverse.  Bonsack's
plot of $H_{\rm eff}$ folded on the 3-day orbital
period appears random.  These results are consistent
with nearly magnetic and rotational pole-on observations.
Bonsack sought other periods, and found marginal
evidence for a 15.88-day period.
He concluded $<H_{\rm eff}>\approx -700$\,G,
with an rms error of a single observation of 140\,G.

\begin{figure}
\includegraphics[width=55mm,height=84mm,angle=270]{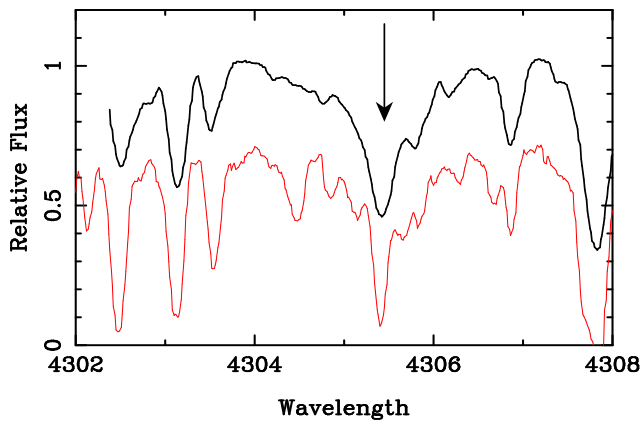}
 \caption{The subordinate line Sr\,{\sc ii} at 4305.47 \AA.
The transition is 5p$^2$P$^o_{3/2}$ -- 6s$^2$S$_{1/2}$.
The lower excitation potential is 3.04 eV.  The upper
spectrum is of HR 710, while the lower is of $\gamma$ Equ.
The latter spectrum is displaced downward by 0.3 for
display purposes.
Although both stars are classified as strontium stars,
the wings on $\lambda$4305 are arguably more pronounced
in HR 710 than in the sharper-lined spectrum of $\gamma$ Equ.
}
\label{fig:4305}
\end{figure}

One of us, CRC, measured wavelengths of a 2.4\,A\,mm$^{-1}$
spectrum from the Dominion Astrophysical Observatory in 1979.
Contemporary notes include a remarks on the strengths
of Sr\,{\sc ii} and Ca\,{\sc ii} K.  Additionally, they note
the subordinate
Sr\,{\sc ii} line, 4305.45 ``has huge wings,''
(see Fig.~\ref{fig:4305}).
Our description is similar to that of Babcock (1958), who
also pointed out that {\it Lines of Sr\,{\sc ii}, like the K line,
have strong, sharp cores with wide shallow wings.}  Much
of the present study follows this lead.

Otherwise, the spectrum
was not found unusual among the variegated spectra
of CP types.  Iron-group spectra, including Sc\,{\sc ii} (but
not V\,{\sc ii}), are rich, while Y\,{\sc ii} and Zr\,{\sc ii} are not as
strong as one might expect given the outstanding strength
of Sr\,{\sc ii}.  The lanthanides are present, but not remarkable
richness or in strength.  The strong Eu\,{\sc ii} lines at
$\lambda\lambda$4129 and 4205, for example, are
exceeded in strength by nearby lines of iron-peak
elements.


We know of no modern abundance study of HR 710.  It was
among several Ap stars analyzed in the 1960's,
with rather crude
methods by today's standards, and only selected elements
were studied.  For example, the work of Searle, Lungershausen,
and Sargent (1966) was restricted to Ti, Mn, Cr, and Fe.
We would agree with these authors that these particular
elements are all enhanced by similar factors.

HR 710 was among the stars studied by Cowley, et al.
(2007, henceforth, CHCGW) for the calcium isotopic anomaly
discovered by Castelli and Hubrig (2004).  In the course of
this work we confirmed  that conventional models failed
to account for the profiles of the strong Ca\,{\sc ii} lines.
This was in complete
agreement with results of Ryabchikova and her
coworkers.  See, for example,
Ryabchikova, et al. (2002, Fig. 3).  These
workers have used a stratified-abundance model
to fit  Ca\,{\sc i} and {\sc ii} and other metallic lines, as
described in subsequent studies on numerous magnetic
Ap stars (cf. Ryabchikova, Kochukhov, and Bagnulo 2007).

For $\gamma$ Equ,
we found that essentially the same model fit the
$\lambda$8542 line of the infrared triplet (henceforth, IRT,
cf. Fig. 7 of CHCGW).  The same stratification
provided a reasonably good fit to the $\lambda$8498-line.

In the course of research leading to CHCGW a number
of lines were synthesized that were not discussed in
that work.
Provisional calculations for HR 710 showed
we would have difficulty
matching the profiles of the IRT with the same stratification
that fit the Ca\,{\sc ii} K-line.

The unusual strength of Sr\,{\sc ii} in HR 710 deserves
closer study.  Would
a stratified model be required to match the profiles
of the Sr\,{\sc ii} resonance lines $\lambda\lambda$4077 and 4215,
or the strong subordinate lines?  If so, would the study
of these features add to our understanding of the
structure of the photospheres of magnetic stars?
The sections below deal with these questions.

We can place only loose constraints on the
physical conditions in the atmosphere of HR 710.
In addition to line core/wing and ionization and excitation
anomalies, there is evidence for abundance patches
on the surface, which can be seen in some of the line
profiles.

\section{Observations}
\label{sec:obs}

The ESO spectra used in this study were obtained on 19 September
2005 with
the UV-Visual Echelle Spectrograph (UVES) at UT2 of the VLT.
The spectra used the new (November 2004)
standard Dichroic settings were described by CHCGW,
covering 3290 to 9460\,\AA.
This configuration makes it possible to measure the $\lambda$8542-line,
which had previously been unavailable because of an order gap.
The resolving power is 80\,000 in the blue-violet, and 110\,000
in the red.  The S/N ratios of the spectra differ, but range
from about 100 to 400.

\section{Reductions}

The UVES spectra were reduced by the UVES pipeline Data Reduction
Software (version 2.5), which is an
evolved version of the {\sc echelle} context of {\sc midas}.
The spectra were mildly Fourier filtered
before the wavelengths were measured at Michigan.

We choose the
continuum by picking high points of the relative flux using
console displays 40~\AA\, wide.  In some cases, a Paschen
line or water-vapor band would
take up most of the display, and no continuum point would
be chosen.  The high points were connected by a spline
{\it and} by linear interpolation.  After some experimentation
the linear fit was given weight 2, the spline 1, and
the weighted average adopted.
The continuum
height is surely uncertain to a few per cent.  Nevertheless
we note that
in the illustrations that follow, no additional (ad hoc)
adjustments were made to the continua.

Wavelength measurements from 4527 to 4652~\AA\, are
from 2.4 A\,mm$^{-1}$ spectra
(12766U, and V)
obtained at the Dominion Astrophysical Observatory in 1979.
The instrument has been described by Fletcher et al. (1980).

\section{The atmosphere of HR 710}
\subsection{Preliminary remarks}

Ap stars were once thought to have abnormal abundances,
but normal atmospheres.  We now know that this is
not the case.  The profiles of the
low Balmer and the strong
metallic lines cannot be reproduced by standard
plane-parallel atmospheres.   Simple adjustments to
temperature, gravity, and composition do not allow one to
resolve inconsistencies in ionization and excitation.
These difficulties have become apparent relatively recently.
Horizontal chemical inhomogeneities (abundance
patches) have been known
in magnetic stars for many decades.

Extensive observations by Babcock (1958) and
Bonsack (1981) show that
spectral variations in HR 710 are mild or absent.
Nevertheless,  abundance patches
apparently influence the spectrum.
We acknowledge
this, among other complications, but do not treat it
in the present study.

\subsection{Photometry}

We used several codes based on Str\"{o}mgren or Geneva photometry
to try to fix $T_{eff}$ and $\log{g}$.  In addition, we
considered ionization balance, Balmer, and Paschen
profiles.  The various implementations give a fairly wide
range of values for $T_{eff}$ (8400 to 8600\,K) and
$\log{g}$ (3.0 to 4.0).

An implementation of Geneva
Photometry by Kaiser (2006), based on K\"{u}nzli, et al.
(1997) includes an
adjustment for [Fe/H]. The
code gives values of $\log{g}$ that decrease with increasing
[Fe/H].  Allowed entries for [Fe/H] are $-1$, 0, and +1.
The relation for both $\log{g}$ and $T_{eff}$ is very nearly
linear.  We used
photometric measurements for
HR 710 from the Simbad data base
to obtain second-order fits, with very
small quadratic terms.   With
[Fe/H] = +1.0, we obtain
$T_{eff} = 8460$\,K and $\log{g} = 3.6$.  This somewhat
low gravity is supported by our ionization balance
of neutral and singly ionized Fe,
where we get best agreement with $\log{g} = 3.0$.
This ionization balance (see below) gives a temperature nearly
200\,K higher.

Codes based
on Str\"{o}mgren photometry give significantly
higher gravities, though they do not,
to our knowledge, include a
provision to allow for metallicity.

The higher surface gravities
agree with Hubrig, et al. (2007), who give
$\log{g} = 4.2$.  The temperature estimate from
this study is 8433\,K.

\subsection{Balmer lines}

Hydrogen line strengths depend sensitively on temperature
in cooler stars, and on surface gravity in hotter ones.
HR 710 is at the high-temperature end of the region where
the line strengths change from primarily temperature to
primarily gravity sensitivity.

All Balmer calculations are based on the profiles
from Stehl\'{e} \& Hutcheon (1999).

\begin{figure}
\includegraphics[width=55mm,height=84mm,angle=270]{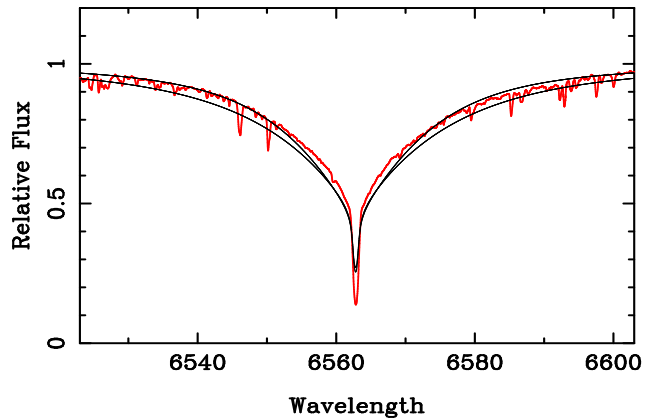}
 \caption{H$\alpha$ profile in HR 710.  The calculated profiles
from models with $\log{g} = 3.0$ (upper curve), and 4.0 (lower
curve).  In both cases, the assumed temperature is 8600\,K.
}
\label{fig:alpha}
\end{figure}

Fig.~\ref{fig:alpha} shows the observed and calculated
profile of H$\alpha$ for models with $T_{eff} = 8600$\,K,
and $\log{g} = 3.0$ and 4.0.
The fit is arguably better
with the lower-gravity model (upper smooth curve).
Metallic lines have been omitted from the calculation
to avoid confusion.  The tendency of the lower Balmer
cores of cooler magnetic Ap stars to be deeper, and
sharper than the observations was called the
core-wing anomaly (Cowley, et al. 2001).  LTE calculations
of the H$\alpha$ core are never as deep as the observations.
In the case of HR 710, departures of the calculations
from the observed profiles in the regions outside the
inner core are of interest.  For H$\alpha$ and especially
for H$\gamma$ and H$\delta$ the calculations do not fit
the observations until the far wings are reached.

\begin{figure}
\includegraphics[width=55mm,height=84mm,angle=270]{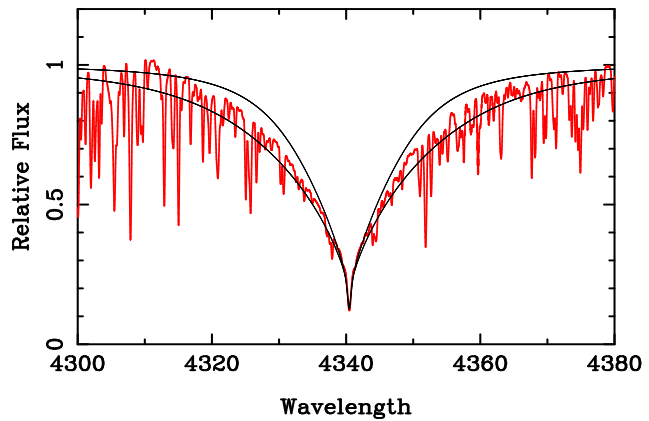}
 \caption{H$\gamma$ profile in HR 710.  The calculated profiles
from models with $\log{g} = 3.0$ (upper curve), and 4.0 (lower
curve).  In both cases, the assumed temperature is 8600\,K.
}
\label{fig:gamma}
\end{figure}

The case for the lower gravity is somewhat stronger with
H$\gamma$ (Fig.~\ref{fig:gamma}) and H$\delta$ (not shown).

\begin{figure}
\includegraphics[width=95mm,height=75mm,angle=0]{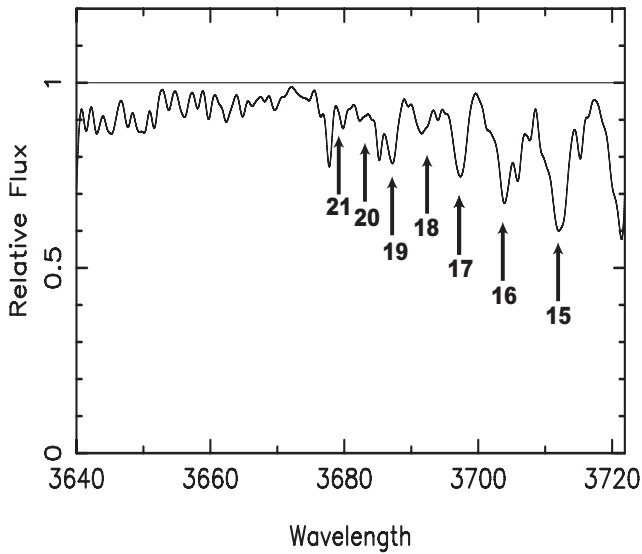}
 \caption{Region of the Balmer confluence.  The spectrum
has been rather severely filtered in an attempt to make the
Balmer lines stand out.
}
\label{fig:balcon}
\end{figure}

Fig.~\ref{fig:balcon} shows the region of the Balmer
confluence.  H18 (3691.56) is surely present, and
H19 (3686.83) possibly so,
though the feature is clearly strengthened by blending
primarily with Cr\,{\sc ii} 3687.03 and Fe I-75 3687.10.  The
Multiplet Number (75) is from
Moore(1945).  The
strong Cr\,{\sc ii} line does not appear in the Multiplet Tables.
The Inglis-Teller relation (cf. Cowley 1970, Eq. AI-5)
yields $\log{N_e} = 13.4$ if the last Balmer line is
H19.  This electron density is reached in our models
with $T_{eff} = 8600$\,K, and
with $\log{g} = 4$ at a logarithmic optical depth
$\log{\tau_{3640}} = -0.7$.  For the $\log{g} = 3$
model the corresponding optical depth is $-0.08$.

We have used the optical depth at $\lambda$3640,
short of the Balmer Jump.
This should closely approximate
the opacity where the confluence occurs, {\it including}
the opacity due to overlapping higher Balmer lines.
The basis of this statement may be found,
among other places, in Cowley (2000, see also
Appendix A).

We expect the electron density obtained
by the Inglis-Teller formula would correspond to
$N_e$ at optical depth unity for the wavelength of
the Balmer jump.  Thus,
the last countable Balmer line appears to give
mild support for a lower-$\log{g}$ model.  However
a plot similar to Fig.~\ref{fig:balcon}
for $\gamma$ Equ shows
Balmer components possibly as high as H20.
Moreover, in two
other main-sequence stars,
$\nu$ Cap (B9.5 V) and $\sigma$ Boo (F2 V), we can
count to H22 or H23 and H19 or 20 respectively.
A
detailed study of the usefulness of the last
countable Balmer line as an indicator of surface
gravity would be worth while.

\subsection{Paschen lines}

The lower Paschen members are beyond our wavelength
coverage.  The first usable line is $\lambda$9229,
P9.  The lowest Paschen members, P9, P10, and P11,
are clearly fit better with a $\log{g} = 4$ model
than $\log{g} = 3$.  All Paschen line calculations
used Lemke's (1997) revised VCS profiles (Vidal,
Cooper \& Smith 1973), because the Stehl\'{e}-Hutcheon
tables do not extend to large enough values of $N_e$
for the higher Paschen lines.

\begin{figure}
\includegraphics[width=55mm,height=84mm,angle=270]{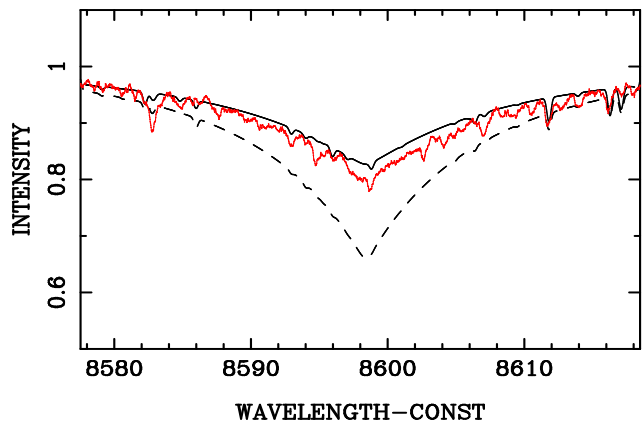}
 \caption{The Paschen line P14.  The smooth, solid curve was
calculated using a model with $T_{eff} = 8600$, and
$\log{g} = 4$.  The dashed line shows the profile from
the lower-gravity model, $\log{g} = 3$.
}
\label{fig:8598}
\end{figure}

At $\lambda$8750, P12, the
lower-gravity model provides a better fit to the wings
than the $\log{g} = 4$ model, but
the core is too deep.  If we were to lower the observed
continuum by 2 to 3\%, the overall fit, core and wings
is significantly better with the higher-gravity model.
Fig.~\ref{fig:8598} shows the Paschen line at $\lambda$8598,
P14, which falls between the longer-wavelength pair of the
IRT.  Again, the higher-gravity model provides a better
fit.

One may surely count Paschen lines to P18.  It
is arguable whether higher Paschen members are present.
Thus the limiting Balmer and Paschen series are in
agreement with one another, but a definitive interpretation
awaits further study.  Better observational material,
specifically taken and reduced for the purpose of
investigation of series confluence would also be
useful.

The Paschen lines P13, P14, P15, and P16 are blended with
present special problems because
they are near the series confluence.  Details are
explained in Appendix A.

\section{Ionization equilibrium}

Lines of Fe\,{\sc i} and {\sc ii} were chosen from a compilation of
oscillator strengths by Fuhr and Wiese (2006).  We
used our line identification list (see URL below) to
exclude cases with obvious blends.  We then calculated
abundances for models with $\log{g} = 3$ and 4 for $T_{eff}$
ranging from 8000 to 8800\,K.  Calculations for all of these
cases were made with, and without stratification.  The
stratification assumed was based on an analytic formula:

\begin{equation}
g(x) = b + (1-b)\left[\frac{1}{2}\sqrt{\pi/a}\pm
\frac{1}{2}{\rm erf}(\sqrt{a\cdot |x+d|^2}\right].
\end{equation}

In any given layer, the abundance of an element is
multiplied by this $g(x)$.
Here, $x = \log{(\tau_{5000})}$, and
the negative sign is taken for $x \le -d$.
This function is a smoothed step of height
$1-b$, with the center of the jump located at depth
$\log{\tau_{\rm 5000} = -d}$.  At large depths, $g(x)$
is nearly unity, while for the smallest depths, it
has the value $b$.
The Gaussian $1/{\rm e}$
width of the step is $1/\sqrt{a}$.

For all spectra {\it other than those of Ca\,{\sc ii} and Sr\,{\sc ii}},
``standard parameters'' were used: $a = 6.7$, $\log{b} = -4$,
and $d = 1$.  We cannot claim any special validity for these
particular values; they were used in CHCGW to reconcile
profiles of the Ca K-line and one line of the IRT.  We have
already explored a wide domain of parameter space, and must
leave possible adjustments of the stratification parameters
to account for anomalous ionization and excitation to
future work.

Tables of wavelengths and equivalent widths may be found
on our web page:
\newline
\verb=http://www.astro.lsa.umich.edu/~cowley/hr710=
\newline For iron, abundances were found for 5
temperatures and 2 gravities, above and below the Balmer
Jump, and
with and without stratification
using the standard parameters.  We made plots of abundances
versus equivalent widths and excitation potential,  and used
the former plots to fix the microturbulent parameter.
Trends for iron were removed by the value
$\xi_t = 3.5$.  This value is somewhat large, but it
also must account for the Zeeman broadening, which was
not large enough to yield resolved patterns.

\begin{figure}
\includegraphics[width=55mm,height=84mm,angle=270]{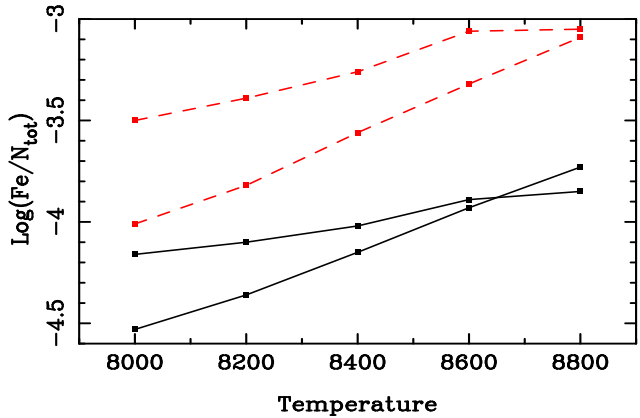}
 \caption{Ionization equilibrium: Fe\,{\sc i} and {\sc ii}.  The dashed
lines are Fe\,{\sc i} (upper) and Fe\,{\sc ii} (lower) using the standard
stratification parameters.  The lower lines are Fe I
(mostly upper)
and Fe\,{\sc ii} (mostly lower) without stratification.
Values of log(Fe/N$_{\rm tot}$) refer to the deep photosphere
(prior to the application of the function $g(x)$).
The four curves are based on lines with wavelengths
longer than 3650~\AA.
}
\label{fig:ironeq}
\end{figure}

The results are illustrated
in Fig.~\ref{fig:ironeq}.  Here, and elsewhere in this paper,
 `$\rm N_{\rm tot}$'  means the sum of relative abundances of
all elements.  If log(H) = 12, then
$\rm N_{\rm tot} = 1.03\cdot 10^{12}$.
We used 70 lines of Fe\,{\sc i} and
32 lines of Fe\,{\sc ii}, all with wavelengths longer than 3650~\AA.
This plot is the primary basis
for consideration of low-gravity (or low pressure) models.

For the 8600\,K, $\log{g} = 3$ model, with no stratification,
we find
\begin{displaymath}
\log{\rm (Fe/N_{\rm tot})} = -3.93\,({\rm Fe\,I})
;\hspace{1mm} \log{\rm (Fe/N_{\rm tot})} = -3.89\,({\rm Fe\,II})
\end{displaymath}

For iron, we made separate abundance determinations for lines
below the Balmer Jump.  For 43 Fe\,{\sc i} lines, we obtained
$\rm \log(Fe/N_{\rm tot}) = -3.86$.  This is less than 0.1 dex different
from the value found for lines above the Balmer Jump.
Only 8 lines of Fe\,{\sc ii} were found usable below the Balmer Jump.
These yielded $\rm \log(Fe/N_{\rm tot}) = -4.02$.  In view of the
smaller number of lines, we do not consider this
value significantly different from the others.
All calculations of iron below the Balmer Jump
were made without assuming stratification.  We
see no compelling evidence for stratification in iron.

The next richest spectra to Fe\,{\sc i} and {\sc ii} belong to Cr\,{\sc i} and {\sc ii}.
We analyzed 36 Cr\,{\sc i} lines and 47 Cr\,{\sc ii} lines.  Oscillator strengths
were taken from VALD (see Kupka, et al.  1999), but we
omitted lines from intersystem transitions, where the
multiplicity of the upper and lower terms are different.
Experience has shown that {\it ab initio} and
semi-empirical calculations for
LS-permitted transitions
are somewhat more reliable than those for intersystem lines.
Reasonable agreement was found for abundances from
the Cr\,{\sc i} and {\sc ii} lines with wavelengths longer than
3650~\AA, from the 8600\,K,
$\log{g} = 3$ model, with no stratification:

\begin{displaymath}
\log{\rm (Cr/N_{\rm tot})} = -4.00\,({\rm Cr\,I)};
\hspace{1mm} \log{\rm (Cr/N_{\rm tot})} = -4.14\,({\rm Cr\,II)}
\end{displaymath}

Several other elements are observed in two stages of
ionization, but our results from them were less concordant.
In particular, the spectra of Ti\,{\sc i} and {\sc ii} may be brought
into better agreement by adjusting model parameters {\it or
introducing stratification}.
Details are tedious, and
are omitted, pending a later investigation.

\section{Calcium}
\subsection{The weaker Ca\,{\sc i} and {\sc ii} lines}

We measured equivalent widths of 23 lines of Ca I, all
with wavelengths longer than the Balmer Jump.  The
equivalent widths spread evenly between 10 and 112 m\AA.
With the 8600\,K, $\log{g} = 3$ model, we obtained
a mean log(Ca/N$_{\rm tot}$) of $-$4.65 $\pm 0.08$
($\xi_t = 3.5$).
There is no significant trend of the
abundance with either equivalent width or excitation
potential; the latter span a narrow range of
1.88 to 2.93 eV.

The Ca\,{\sc i} resonance line, $\lambda$4227 was not included
in the above average.  It is only moderately strong
($W_\lambda = 169$ mA), and its core is in not at all
unusual, like those of the K-line or
$\lambda\lambda$4077 or 4215 of Sr\,{\sc ii}.  We can fit the
profile with log(Ca/N$_{\rm tot}$) of $-$4.65, and
lowering the $\xi_t$ to 1.5 km\,s$^{-1}$.  We conclude
$\lambda$4227 is in reasonable agreement with the weaker
Ca\,{\sc i} lines.

We found only 7 useful Ca\,{\sc ii} lines, exclusive of
H and K, and the IRT.
The mean value of log(Ca/N$_{\rm tot}$) from all 7 lines
was $-$4.99 $\pm 0.33$, with the 8600\,K, $\log{g} = 3$ model.
With these lines there were systematic trends both
with excitation potential and equivalent width.
Four lines, with excitation potentials averaging
7.2 eV gave log(Ca/N$_{\rm tot}$) = $-$5.5, while three
lines with average $\chi_{\rm ex}$ of 9.0 eV gave
log(Ca/N$_{\rm tot}$) = $-$4.2.  The tendency of high-excitation
lines to give anomalously high abundance is a characteristic
of the newly discussed anomalies of CP stars
(cf. Ryabchikova, et al. 2004).  However,
the abundance averages from
Ca\,{\sc i} and {\sc ii} are within the standard errors; we have
not attempted to deal with the systematic trends.

One Ca\,{\sc ii} line is below the Balmer Jump:
3461.87 ($\chi = 7.51$).  The abundance from this
single line is log(Ca/N$_{\rm tot}$) = $-$4.6.  We
make no judgement about stratification from this
single line.

\subsection{The K-line}

\begin{figure}
\includegraphics[width=55mm,height=84mm,angle=270]{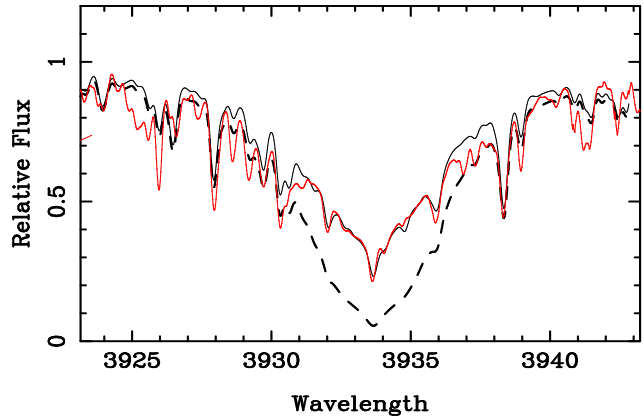}
 \caption{The Ca\,{\sc ii} K-line, observed (solid gray) and
calculated with stratification.  The stratification
parameters for $g(x)$ are: $a = 6.7$, $\log{b} = -7.0$,
$d = 0.25$, and $\rm \log{(Ca/N_{tot})} = -3.10$.
The
heavy dashed line is for $\rm \log{(Ca/N_{tot})} = -4.65$,
and {\it no} stratification.
}
\label{fig:CaK}
\end{figure}

The Ca\,{\sc ii} K-line exhibits the full symptoms of stratification.
An attempt to fit it with the abundance obtained from the
Ca\,{\sc i} or Ca\,{\sc ii} lines (cf. above) results in a deep, broad
core, as shown in Fig.~\ref{fig:CaK}.  The plot is
quite similar to Fig. 3 of Ryabchikova, et al. (2002)
for the star $\gamma$ Equ.  Here, and in similar figures
to follow, only token efforts were made to fit the
atomic line spectrum apart from the feature of direct
interest.

The stratification parameters given in the caption to
Fig.~\ref{fig:CaK} were determined by trial and
error.  They imply an abundance jump of 7 orders of
magnitude between optical depths
$\log{(\tau_{5000})} = -0.7$ and +0.3.

\subsection{The IRT}

Lines of the IRT manifest stratification in a way that
is somewhat similar to the Ca\,{\sc ii} K-line, though the
total absorption is much less, and the central absorption
is more significant relative to the total profile.
We must keep in mind that the possible influence
of surficial abundance patches which could
have an influence on the profiles of the IRT.
In view of the better fits with
$\log{g} = 4$ for P14, which falls between
$\lambda$8498 and $\lambda$8542 of the IRT, as well as
lowerPaschen all calculations illustrated for the IRT
have used the higher gravity.

\begin{figure}
\includegraphics[width=55mm,height=84mm,angle=270]{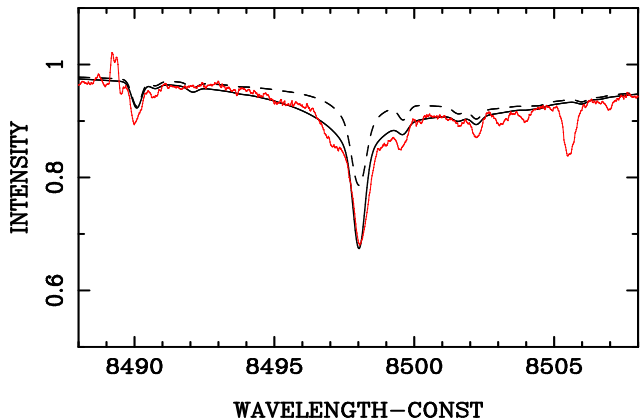}
 \caption{The Ca\,{\sc ii} $\lambda$8498-line, observed (solid gray) and
calculated (solid black) with stratification.  The stratification
parameters for $g(x)$ are: $a = 6.7$, $\log{b} = -5.70$,
$d = 0.25$.  $\rm \log(Ca/N_{tot}) = -2.12$.
The
dashed line is for the parameters used to fit the Ca\,{\sc ii} K-line
(see Fig.~\ref{fig:CaK}).  Both calculations employed a  model
with $\log{g} = 4$.
}
\label{fig:8498}
\end{figure}

Fig~\ref{fig:8498} shows observed and computed profiles
for $\lambda$8498 in HR 710.  Damping wings are present,
but the stellar profiles change rapidly in structure, from
core to wing.  These shapes can arise if the abundance
of calcium varies with a stratification function $g(x)$
as described above.

Note the extraordinarily high value of calcium required for
the deep photosphere, and the large abundance gradient
implied by the parameters of the function $g(x)$.

\begin{figure}
\includegraphics[width=55mm,height=84mm,angle=270]{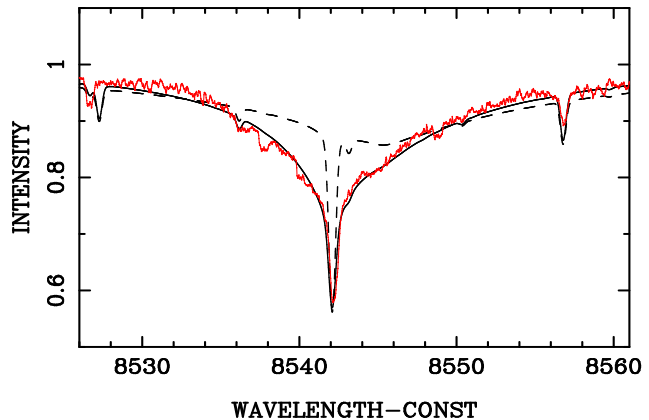}
\caption{The Ca\,{\sc ii} $\lambda$8542-line, observed (solid gray) and
calculated (solid black) with stratification.  The stratification
parameters for $g(x)$ are: $a = 1.0$, $\log{b} = -6.70$,
$d = -1.0$.  $\rm \log(Ca/N_{tot} = -1.52$.
The function $g(x)$
differs from unity throughout the atmosphere, so the effective
$\rm \log(Ca/N_{tot})$ at $\log{(\tau_{5000})} = 0.0$ is roughly
$-2.52$,
in fair agreement with the value necessary
to fit the $\lambda$8498 line.  The
dashed line is for the parameters used to fit the
$\lambda$8498 line.  Both calculations use
$\log{g} = 4$.
}
\label{fig:8542}
\end{figure}

The fit to the strongest line of the IRT, $\lambda$8542, is
shown in Fig.~\ref{fig:8542}.  The wings of this line are
significantly stronger than those of the weaker $\lambda$8498
line.
The best fit (solid line) used different
stratification
parameters from those used to fit the weaker $\lambda$8498
line (dashed line), though the abundance in the deep photosphere is
roughly similar: $\rm \log(Ca/N_{tot}) = -2.5$ vs. $-2.1$.
Several lines of neutral calcium are in the region of
Fig.~\ref{fig:8542}.  Two of the lines,
at 8525.72 and 8555.51, were strong enough
to require explicit inclusion
in the calculation.  The scaling factor $g(x)$ had
to be applied to them, as well as $\lambda$8542.  Otherwise,
with the Ca/N$_{\rm tot}$ assumed for the deep photosphere,
they would have been huge.

\begin{figure}
\includegraphics[width=55mm,height=84mm,angle=270]{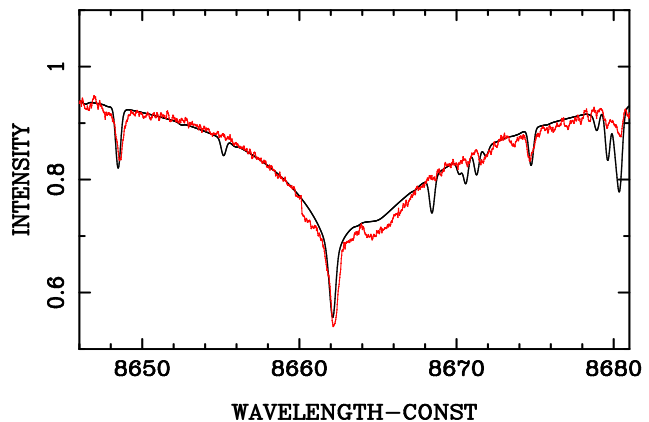}
\caption{The Ca\,{\sc ii} $\lambda$8662-line, observed (solid gray) and
calculated (solid black) with stratification.  The calcium
abundance and stratification parameters are {\it identical}
to those used to fit the $\lambda$8542-line.  The model has
$\log g = 4$.
}
\label{fig:8662}
\end{figure}

Fig.~\ref{fig:8662} shows the third member of the IRT,
$\lambda$8662.  No
adjustment of the parameters from those used to fit
the $\lambda$8542 line has been made.

\section{Strontium}
\subsection{Weaker subordinate lines}

In addition to the resonance lines $\lambda\lambda$4077 and 4215,
and the stronger subordinate lines $\lambda\lambda$4161 and 4305,
we found five marginally unblended lines of Sr\,{\sc i} and {\sc ii}, and
determined provisional abundances from them.  Results are shown
in Table~\ref{tab:weaksr}
for the 8600\,K, $\log(g)=3$ model.
The abundances are up to a factor of 3 lower if the
$\log(g) = 4$ model is used, but the agreement between the
first and second spectra is still within 0.1 dex for the lower
microturbulence.  The weaker, Sr\,{\sc i} lines, which are less sensitive
to microturbulence, magnetic broadening, and surface gravity
give mean abundances of $-$6.3 and $-$6.4 for the lower and
higher values of $\xi_t$ respectively.

We have not experimented
extensively with stratification.  The parameters used to
get a fit for $\lambda$4077 (see below) give unacceptably high
abundances from the weak Sr\,{\sc i} lines.  The ``standard'' stratification
parameters, $a = 6.7$, $\log{b} = -4.0$, and $d = +1$,
reduces the abundance difference between the first and second
spectra though only by a little, and the overall abundances
increase by more than a dex.

We conclude that a reasonable mean abundance for the atmosphere
of HR 710 is $\rm \log(Sr/N_{\rm tot}) \approx -6.4$.  This is
a factor of 2.5 higher than the abundance found by
Ryabchikova, et al. (1997) for the strontium star $\gamma$ Equ.
It comports well with our assertion, based on Fig.~\ref{fig:4305},
that strontium is somewhat stronger in HR 710 than
$\gamma$ Equ.

\begin{table}
\centering
  \caption{Abundances from weaker subordinate Sr lines: 8600\,K,
  $\log(g) = 3$.\label{tab:weaksr}}
  \begin{tabular}{l c c c c}
  \hline
Spectrum &$\lambda$[\AA]&$W_\lambda$[mA]
         &\multicolumn{2}{c}{$\rm\log{(Sr/N_{\rm tot})}$} \\
         &              &                &   $\xi_t = 1.3$   &$\xi_t = 3.5$   \\ \hline
Sr\,{\sc ii}    &3380.69       & 97             &$-$6.39&$-$7.70  \\
Sr\,{\sc ii}    &3474.89       & 83             &$-$5.96&$-$7.23  \\
Sr\,{\sc i}    &6408.44       & 23              &$-$6.21&$-$6.31  \\
Sr\,{\sc i}    &6504.00       & 20              &$-$6.07&$-$6.15  \\
Sr\,{\sc i}    &6617.27       & 18              &$-$5.87&$-$5.93  \\
Averages       &              &                 &$-$6.10&$-$6.66 \\
log(Sr/Sr$_\odot$)&              &              & +3.02 & +2.96\\ \hline
\end{tabular}
\end{table}

\subsection{The resonance lines}

\begin{figure}
\includegraphics[width=55mm,height=84mm,angle=270]{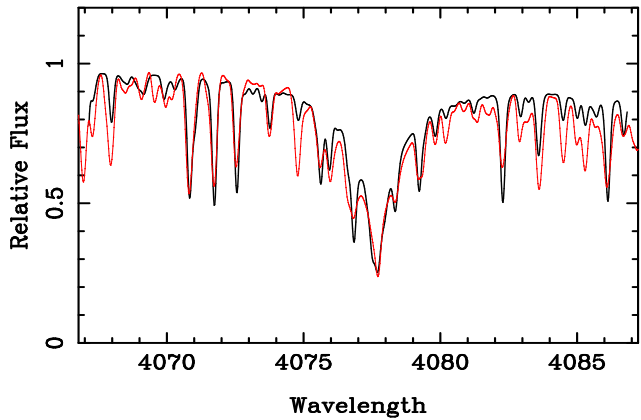}
\caption{The Sr\,{\sc ii} $\lambda$4077-line, observed (solid gray) and
calculated (solid black) with stratification.  The
parameters of the fit are: $a = 1.7$, $\log{b} = -6.70$,
$d = -1$, $\rm \log(Sr/N_{tot}) = -4.00$.
At
$\log{(\tau_{5000})} = 0.0$, $\rm \log(Sr/N_{tot}) = -4.85$.
This is still considerably larger than the value obtained
from the weak lines (Table~\ref{tab:weaksr}):
$-6.10$, with $\xi_t = 1.3$.
}
\label{fig:4077}
\end{figure}

We were unable to fit the near wings of the stronger
resonance line, $\lambda$4077, without moving the
abundance jump deep into the photosphere.  The
adopted parameters given in Fig.~\ref{fig:4077}
lead to $g(x) = 0.770$ at the deepest layer
of our atmosphere, $\log{(\tau_{5000})} = 1.4$.
The strontium fraction decreases rapidly from
the deepest layers, and levels to the value
$b\cdot$\rm (Sr/N$_{tot})$ = $2\cdot 10^{-11}$ by
$\log{(\tau_{5000})} = -5.4$

\begin{figure}
\includegraphics[width=55mm,height=84mm,angle=270]{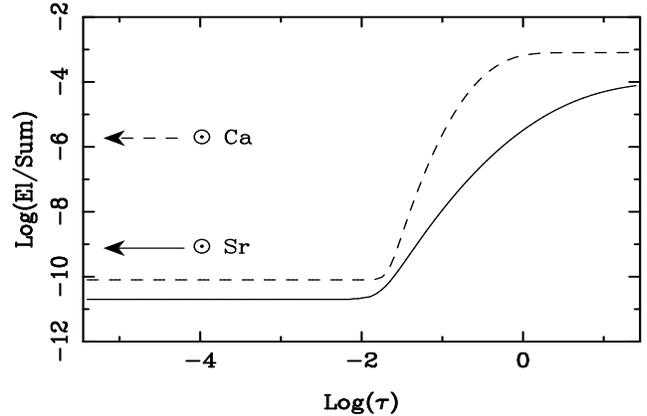}
\caption{Stratified abundances of strontium (solid) and
calcium (dashed) using parameters appropriate for the
$\lambda$4077 and K-line, respectively.  Solar abundances
are indicated.
}
\label{fig:strat}
\end{figure}

Fig.~\ref{fig:strat} shows the abundance stratification
profiles of strontium and calcium.

\begin{figure}
\includegraphics[width=55mm,height=84mm,angle=270]{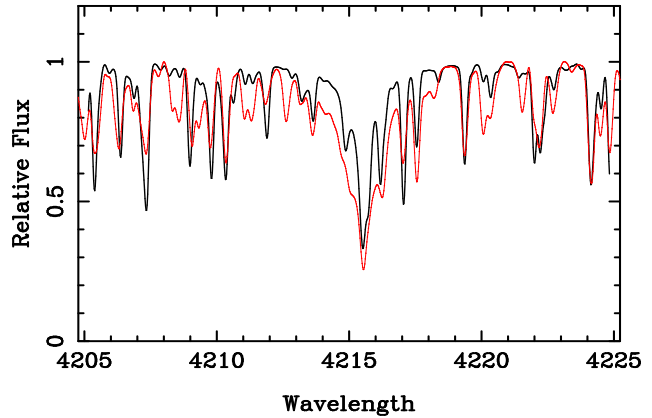}
\caption{The Sr\,{\sc ii} $\lambda$4215-line, observed (solid gray) and
calculated (solid black) with stratification.  The
parameters {\it are the same} as those used in the fit
for $\lambda$4077.  A considerably better fit
(not shown) can be
obtained from a slight modification of the parameters
to: $a = 1.7$, $\log{b} = -6.70$,
$d = -1$, $\rm \log(Sr/N_{tot}) = -3.70$.
}
\label{fig:4215}
\end{figure}

The weaker of the resonance lines, $\lambda$4215, is shown
in Fig.~\ref{fig:4215}.  The calculation shows used the
same fitting parameters
as those used for the stronger line.
It is unclear just how meaningful this might be, since the
blending is quite significant, and extends well into the
core of the profile.

\subsection{The subordinate Sr\,{\sc ii} lines}

\begin{figure}
\includegraphics[width=85mm,height=74mm,angle=0]{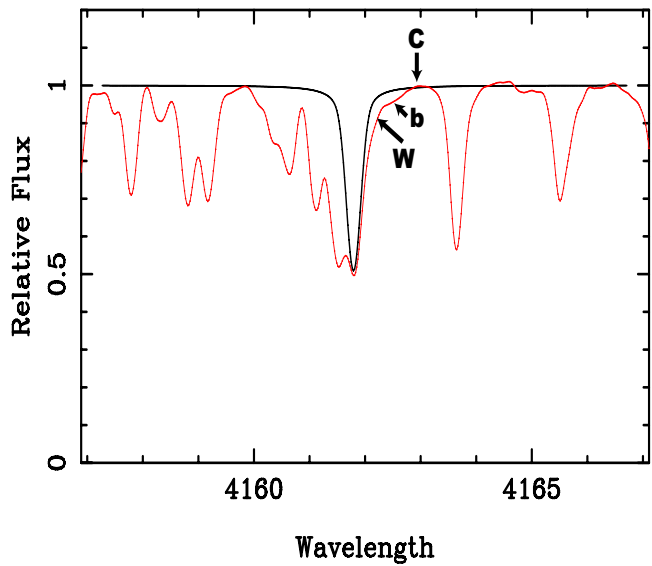}
\caption{The Sr\,{\sc ii} $\lambda$4162-line, observed (solid gray) and
calculated (solid black) with stratification.  The
parameters of the fit are: $a = 1.7$, $\log{b} = -7.0$,
$d = -1$,  $\rm \log(Sr/N_{tot}) = -3.70$,
the same as for the
$\lambda$4077 resonance line.  Other atomic lines have been
omitted in the calculated profile for clarity.  The near
wing (W) and continuum (C) are noted.  The observed violet
wing contains a weak absorption, primarily due to Fe\,{\sc i} (b).
It is not strong enough to confuse the location of the near
wing of the Sr\,{\sc ii} line.
}
\label{fig:4162}
\end{figure}

Fig.~\ref{fig:4162} shows a calculation of the subordinate
line, $\lambda$4161.80, $\rm 5p^2P^o_{1/2}-6s^2S_{1/2}$.
The line is in Sr\,{\sc ii} Multiplet 3.
Only the long-wavelength part of the profile is sufficiently
unblended for use in fixing the stratification parameters.
The calculation shows the result of using the parameters from
the best $\lambda$4215-fit, which is poor.
While we were able to improve the fit in the red wing
somewhat, by trial-and-error adjusting the parameters
of $g(x)$, we never achieved a satisfactory fit.  When
the near wing, marked `W' in the figure, was fit, the
calculated continuum (C) was below the observed one.

A similar
situation was found for the stronger line of Multiplet 3,
$\lambda$4305, though for this line it was the violet wing
that proved most useful (cf. Fig.~\ref{fig:4305}).

In all likelihood, a satisfactory fit could be found with
a more elaborate stratification model than the one fixed
by the parameters of our $g(x)$-function.  We do not
explore this here.

\section{Stratification and HR 710}
\subsection{Summary of stratification parameters}
Stratification parameters are collected in Table~\ref{tab:stratpar}.
The second entries for $\lambda$3933 give a good fit to
the core and wings (to ca. $\pm 40$\,A) for a surface
gravity $\log g = 4.0$.  There is no corresponding figure.
The entries for Sr\,{\sc ii} $\lambda$4215
are the values mentioned in the
caption to Fig.~\ref{fig:4215}, which give an improved
fit.  Note that (El/N$_{\rm tot}$) refers to the deepest
photospheric layers.

\begin{table}
 \centering
  \caption{Summary of stratification parameters\label{tab:stratpar}}
  \begin{tabular}{l c c c c l}
  \hline
$\lambda$[\AA] &log(El/N$_{\rm tot}$)
&$a$&$\log b$&$d$&Rem. \\ \hline
\multicolumn{6}{c}{Ca\,{\sc ii}}  \\ \hline
3933 & $-3.10$&6.7&$-7.00$& 0.25& Fig.~7 \\
3933 & $-4.00$&6.7&$-6.00$& 0.50& log\,$g$\,=\,4 \\
8498 & $-2.12$&6.7&$-5.70$& 0.25& Fig.~8 \\
8542 & $-1.52$&1.0&$-6.70$& -1.0& Fig.~9 \\
8662 & $-1.52$&1.0&$-6.70$& -1.0& Fig.~10 \\ \hline
\multicolumn{6}{c}{Sr\,{\sc ii}} \\ \hline
4077 & $-4.00$&1.7&$-6.70$& -1.0& Fig.~11 \\
4215 & $-3.70$&1.2&$-6.70$& -1.0& See text \\
4162 & $-3.70$&1.7&$-7.00$& -1.0& Fig.~14 \\ \hline
\end{tabular}
\end{table}

\subsection{General remarks}

The near wings and cores of strong lines in magnetic
CP stars cannot
be fit with traditional models.  Non-LTE is surely relevant,
but it is most unlikely to account for all of the discrepancies
between theory and observation.  Modern workers have returned
to the notion of ``anomalous atmospheres, or AA,'' a term
coined by Underhill (1966).  Kochukhov, Bagnulo, \& Barklem
(2002) showed
that the Balmer-line profiles could be fit with an {\it ad hoc}
modification of the temperature distribution in the high
photosphere.  Ryabchikova and colleagues (cf. Ryabchikova 2005)
have used stratified abundance models to explain both the
cores and near-wings of strong lines of Ca\,{\sc ii} {\it and}
anomalous excitation and ionization, often seen in magnetic
CP stars.

One might expect that the same stratification model would fit
all of the Ca\,{\sc ii} lines, H and K, and the IRT.  This was not
the case for HR 710.   We must note that our
stratification model is somewhat more constrained, by
Equ. 1, than those used by Ryabchikova and coworkers.

The cores of the resonance lines of Sr\,{\sc ii},
$\lambda\lambda$4077 and 4215 exhibit similar peculiarities
to the K-line.  Vastly improved fits can be obtained using
stratified models, though slightly different stratifications
are required.  The subordinate lines, $\lambda\lambda$4162 and
4305 require stratification or other modifications to fit
the wings.  The stratification parameters required to
fit $\lambda$4077, however, make very little difference in
the wings of the two subordinate lines.

Ionization and excitation anomalies are present in the HR 710
spectrum.  The Ti\,{\sc i} and {\sc ii} ionization and excitation imbalance
deserves further study.  The rich Fe\,{\sc i} and {\sc ii} spectra were
in satisfactory agreement without the need for stratification,
provided we employed the significantly low-gravity model
with $\log{g} = 3.0$.

Generally, the line spectrum in the red and infrared was fit
better by a higher-gravity model, and the
relevant illustrations shown
are based on that model.  We cannot rule out the relevance of
abundance patches
for our difficulties in fitting
all of the Ca\,{\sc ii} or the Sr\,{\sc ii} lines with a single stratification
model.


\section{HR 710Aa and its companion HR 710Ab}

The mass function for an SB1 system is
$f(m) = m^3\sin^3(i)/(m+M)^2 M_\odot$, where $M$ and $m$
are the masses of the primary and secondary, respectively.
This $f(m)$
was well determined by Bonsack, and has the value
$1.777\cdot 10^{-3}$.

For a given $m$ and $f(m)$, the orbital inclinations
decrease as $M \rightarrow m$.  Kitamura (1980) noted
that the inclinations of short-period Ap SB1 stars tend to be
small.  Such near pole-on views of SB1 systems would
require a subluminous star, since the companion spectrum
is unseen.  Bonsack suggested a white dwarf for HR 710.
This model
explains the failure of the magnetic field to
show an obvious reversal.

If we assume the 3-day orbital period is also the period of
rotation of the primary, the observed $v\cdot\sin{i}$ depends
directly on the radius of the star and
the inclination of its rotational axis.

The equatorial
velocity of a star is
\begin{equation}
v_{\rm eq} = {{50.56\cdot R}\over P},
\end{equation}
\noindent where $R$ is in solar radii and $P$ in days.

Values of $v\cdot\sin{i}$ have been published by Bonsack (1981),
and Preston (1971): 7.7, and 13 km\,s$^{-1}$ respectively.
We have made a new determination from our UVES spectra,
using the magnetic null lines $\lambda\lambda$3850, 4006,
4065, and 7090, all of Fe I.

The observations were
(Fourier) unfiltered, but they were interpolated to
every 0.02\,\AA, for compatibility with Michigan
software.
We assumed a Gaussian 1/$e$-width
of 0.04~\AA\, for the instrumental profile in the blue-violet.
This allows
for a small degredation of the resolution; a resolving power,
${\GR}$ of 80\,000 would give ca. 0.03\,\AA\, for
the 1/$e$-width, where we use
$\Delta\lambda$ = $\lambda/{\GR}$
as the FWHM of the instrumental profile.  For $\lambda$7090,
using ${\GR} = 110\,000$, the 1/$e$ width of the pure
instrumental profile is 0.045\,\AA, and we assumed 0.055\,\AA.

Synthetic spectra were fitted
to regions containing these lines spectra, and the best
fits determined by trial and error.
We find
that a value of 10 km\,s$^{-1}$ gives the best overall
fit.
This is near the mean of the Preston and
Bonsack's published values.  However because of blending,
and subjective judgements about a fit, the uncertainty
of $v\cdot\sin{i}$ is still between 1 and 2 km\,s$^{-1}$.

We prefer $M/M_\odot = 1.84$ and $R/R_\odot = 1.63$ from
tables in Drilling and Landolt (1999), for a main sequence
A-star with an effective temperature of 8600\,K.
The values
agree well with Hubrig, North, and Sch\"{o}ller (2007).

It is now straightforward to get the inclination from
Eq.\,(2), if we assume a 3-day period.
From the inclination,
the secondary mass, $m$, will follow from the mass function,
given a primary mass, $M$.  In order to provide some
indication of the overall uncertainties, we give the
secondary mass in Table~\ref{tab:secondary},
for three assumed masses
(and corresponding radaii), and three values of
$v\cdot\sin{i}$.  We assume the relation between mass
and radius for main sequence stars given by Drilling
and Landolt.

\begin{table}
 \centering
  \caption{Secondary mass $m$ for various primary masses
  $M$ and $v\cdot\sin{i}$.  The preferred
  values are in bold face.\label{tab:secondary}}
  \begin{tabular}{c c c c c c}
  \hline
$M/M_\odot$ &$R/R_\odot$& $i$\,(deg)  &
\multicolumn{3}{c}{$v\cdot\sin{i}$}\\
     &     &   & 8  &{\bf10}& 12  \\ \hline
2.00 &1.74 &29 &0.89&0.68&0.56 \\
{\bf 1.84} &{\bf 1.63} &28 &0.79&{\bf 0.62}&0.48 \\
1.60 &1.46 &25 &0.64&0.49&0.40 \\ \hline
\end{tabular}
\end{table}

Our preferred values give $m = 0.62\,M_\odot$,
conveniently near the
most probable observational value for white dwarf masses
according to Weidemann (1990).  This supports the assertion
of Czesla and Schmitt (2007) that the secondary of HR 710Aa
``is clearly a very plausible candidate for the observed
X-ray emission...".  These authors discuss a magnetically
confined wind-shock model.

Bonsack (1981) also discussed a 15.88-day magnetic period.
With $R = 1.63\,R_\odot$, this period leads to an equatorial
velocity of 5.2\,km\,s$^{-1}$, incompatible with the adopted
$v\cdot\sin{i}$.  To reconcile this period with the observed
$v\cdot\sin{i}$, we would need to assume a stellar radius
nearly double that used.  This is not compatible with the
luminosity of the star.

\section{Remarks and caveats}

The present work on HR 710 shows that it is difficult to
explain the profiles of strong Ca\,{\sc ii} and Sr\,{\sc ii} lines with
a simple abundance-stratified model.  Overall, we
strongly support the stratification models, but find
indications that refinements are necessary.

Different stratification parameters were required for
the K-line and the IRT of Ca\,{\sc ii}, while Sr\,{\sc ii} lines also
required adjustments of the parameters for individual
lines.

It remains to be seen if theoretical work can explain
the large elemental abundances of the best-fit stratification
models. These require very high abundances in the deep
photosphere in order to explain the observed, broad wings.
In the case of calcium, traditional
diffusion calculations have explained low photospheric
abundances (Am stars).

Among a number of uncertainties in the present discussion,
we note that
our stratification model ($g(x)$) may (a) not have
explored the optimum domains of parameter space, or (b)
have been too simplistic.  Regarding the latter point,
we note the sophisticated approach of Kochukhov,
et al. (2006).

Eventually, it
will prove necessary to approach the modeling of the
spectra of these stars with improved methods.
Useful modifications of our procedures might include
changing the $T-\tau$ relation in the high photosphere.
This has already been shown to give fits to the Balmer
lines (Kochukhov, Bagnulo, \& Barklem 2002), and can
give nibs on the K-line (Cowley, Hubrig, \& Kamp 2006).
In the last reference, it was also pointed out that
sharp K-line nibs could result if the gas pressure in
the upper atmosphere were lowered.

Calculations done in LTE must be considered provisional
until supported by work done with non-LTE.  This is
especially true for the upper atmospheres, where the
total radiative flux is very nearly constant,
independent of local conditions.  The core of the Ca K-line
is formed highest of lines considered in this work,
and Cowley, Hubrig \& Kamp showed that LTE calculations
were only slightly modified by non-LTE.

The greatest challenge of the observations of strong-line
profiles of CP stars is to the theoretician.  It is
straightforward to make ad hoc adjustments to a model until a
profile is fit.  It is quite another to justify such
modifications with rigorous calculations.

\section*{Acknowledgments}

We thank Dr. Paul Barklem for permission to use his code
{\sc wcalc}
for calculation occupation probabilities of hydrogen levels,
and Dr. Pierre North, for pointing out the short-period
system HD 200405.
CRC is grateful for past use
of the facilities of the Dominion Astrophysical Observatory.
He also thanks colleagues at Michigan and elsewhere for
helpful comments.  We thank the referee, Lars Freyhammer,
for many valuable comments, corrections, and suggestions.
This research has made use of the SIMBAD database,
operated at CDS, Strasbourg, France.

\section*{REFERENCES}


\hangpar Babcock, H. W., 1958, ApJS, 3, 141.

\hangpar Barklem, P., 2004,
http://www.astro.uu.se/$\sim$barklem/, see hbop.f

\hangpar Bonsack, W. K., 1981, PASP, 93, 756


\hangpar Carrier, F., North, P., Udry, S., Babel, J., 2002,
A\&A, 151, 169.

\hangpar Castelli, F., Hubrig, S., 2004, A\&A, 421, L1


\hangpar Cowley, C. R., 1970, The Theory of Stellar Spectra
(London: Gordon and Breach), p. 213

\hangpar Cowley, C. R., 2000, Obs., 120, 318

\hangpar Cowley, C. R., Hubrig, S., \& Kamp, I. 2006,
ApJS, 163, 393

\hangpar Cowley, C. R., Hubrig, S., Castelli, F., Gonz\'{a}lez,
J. F., and Wolff, B., 2007, MNRAS, 377, 1579 (CHCGW)

\hangpar Cowley, C. R., Hubrig, S., Ryabchikova, T. A., et al.
2001, A\&A, 367, 939

\hangpar Cunto, W., Mendoza, C., Ochsenbein, F., and
Zeippen, C. J., A\&A, 275, L5

\hangpar Czesla, Schmitt, J. H. H. M., 2007, A\&A, 465, 493

\hangpar Drilling, J. S., and Landolt, A. U. 1999, in Allen's
Astrophysical Quantities, 4th ed, ed. A. N. Cox
(Springer/AIP Press), p. 381

\hangpar Fletcher, J. M., Harmer, C. F. W., Harmer, D. L., 1980,
Pub. Dom. Ap. Obs., 15, 405

\hangpar Fuhr, J. R., Wiese, W. L., 2006, J. Phys. Chem.
Ref. Data., 35, 669

\hangpar Hubrig, S., North, P., Sch\"{o}ller, M.,
2007, AN, 328, 475

\hangpar H\"{u}nsch, M., Schmitt, J. H. M. M., Voges, W., 1998,
A\&AS, 132, 155

\hangpar Kaiser, A., 2006, in {\it Astrophysics of Variable
Stars}, ed. C. Sterken and C. Aerts, ASPC, Vol. 349, 257

\hangpar King, J. R., Villareal, A. R., Soderblom, D. R.,
et al. 2003, Astron. J., 125, 1980

\hangpar Kitamura, M., 1980, Astrophys. Sp. Sci., 68, 283

\hangpar Kochukhov, O., Bagnulo, S., Barklem, P. S., 2002,
ApJ, 578, L75

\hangpar Kochukhov, O., Tsymbal, V., Ryabchikova, T.,
Makaganyk, V., Bagnulo, S., A\&A, 2006, 460, 831

\hangpar Kupka, F., Piskunov, N, E., Ryabchikova, T. A.,
Stempels, H. C., Weiss, W. W., 1999, A\&AS, 138, 119

\hangpar K\"{u}nzli, M., North, P., Kuruca, R. L.,
Nicolet, B., 1997, A\&AS, 122, 51

\hangpar Lemke, M. 1997, A\&AS, 122, 285

\hangpar Moore, C. E., 1945, Cont. Princeton. Univ. Obs., 20

\hangpar Preston, G. W. 1971, ApJ, 164, 309

\hangpar Roman, N. G., 1949, ApJ, 110, 205

\hangpar Ryabchikova, T. 2005, in Element Stratification
in Stars: 40 Years of Atomic Diffusion, ed. G. Alecian, O.
Richard, and S. Vauclair, EAS Pub. Ser. 17, 253.


\hangpar Ryabchikova, T., Adelman, S. J., Weiss, W. W.,
and Kuschnig, R. 1997, A\&A, 322, 234

\hangpar Ryabchikova, T., Kochukhov, O., Bagnulo, S., 2007,
arXiv:astro-ph/0703296v1

\hangpar Ryabchikova, T., Nesvacil, N., Weiss, W. W.,
Kochukhov, O., St\"{u}tz, Ch., 2004, A\&A, 423, 705

\hangpar Ryabchikova, T., Piskunov, N., Kochukhov, O.,
Tsymbal, V., Mittermayer, P., Weiss, W. W.,
2002, A\&A, 384, 545

\hangpar Searle, L., Lungershausen, W. T., and Sargent,
W. L. W., 1966, ApJ, 145, 141


\hangpar Stehl\'{e}, C., Hutcheon, R. 1999, A\&AS, 140, 93


\hangpar Sugiura, M. Y., 1927, J. Phys. Radium, 8, 113

\hangpar Tokovinin, A. A., 1997, A\&AS, 121, 71

\hangpar Underhill, A. 1966, in IAU Symp. 26, Abundance
Determinations in Stellar Spectra, ed. H. Hubenet (London:
Academic Press)

\hangpar Weidemann, Y. 1990, Ann. Rev. Astron. Ap., 28, 139

\hangpar Vidal, C. R., Cooper, J., Smith, E. W., 1973, ApJS, 25, 37

\appendix
\section{Calculation of higher Paschen lines}

Calculations near series limits need to account for overlapping
lines, as well as the ``dissolution'' of some discrete
upper levels by various perturbations from the plasma bath
of the emitting and absorbing atoms.  We have implemented
the scheme outlined in Cowley (2000) in conjunction with
a probability function $W(n)$ for the existence of discrete
upper levels.  The specific routine was from
Barklem (2004).
The basic assumption is that we can calculate the total opacity
due to both bound-bound and quasi-continuous absorption by
considering only a wavelength region near a given bound-bound
transition.  These regions extend from the halfway points
between transitions to level $n-1$ and $n$, and from $n$ to
$n+1$.  We shall refer to these regions as the {\it n-local
regions.}  Following the reference cited, we define an
$n$-dependent cross section, such that for a Paschen line
\begin{displaymath}
\int \sigma_{3n} d\Delta\lambda =
\frac{\pi e^2}{mc^2}\lambda_n^2 f_{3n}.
\end{displaymath}
The integral here is assumed to extend {\it only} over the
n-local region.
The cross sections so defined are in excellent agreement
with those found in TOPbase (Cunto, et al. 1993) for
photon energies below the ionization limit.

Line absorption in the n-local region is calculated in the
usual way, but with the product $N_3f_{3n}$ diminished by the
occupation probability $W(n)$.  However, the line profile
is truncated at each end of the n-local region.  The ordinary
continuous opacity is then augmented by an amount necessary
to bring the total absorption into agreement with the above
equation.  In this way the total absorption merges smoothly
with the known photoelectron absorption cross section at
the series limit, as has been known for many years
(Sugiura 1927).

\label{lastpage}
\end{document}